\documentclass[pra,twocolumn,showpacs]{revtex4}

\usepackage{graphicx}
\usepackage{textcomp}
\usepackage{amsmath}
\usepackage[latin1]{inputenc}

\newcommand{\ket}[1]{\ensuremath{\left|#1\right\rangle}}

\begin{document}
\date{\today}
\author{U.~Schnorrberger$^1$}
\author{J.~D.~Thompson$^{1,2}$}
\author{S.~Trotzky$^{1}$}
\author{R.~Pugatch$^{3}$}
\author{N.~Davidson$^{3}$}
\author{S.~Kuhr$^{1}$}
\author{I.~Bloch$^{1,4}$}
 \affiliation{
  $^1$Johannes Gutenberg-Universit\"{a}t, Institut f\"{u}r Physik, Staudingerweg 7, 55128 Mainz, Germany\\
  $^2$Department of Physics, Harvard University, Cambridge, Massachusetts 02138,
  USA\\
  $^3$Department of Physics of  Complex Systems,  Weizmann Institute of Science,
  Rehovot 76100, Israel\\
  $^4$Max-Planck-Institut f\"ur Quantenoptik,
  Hans-Kopfermann-Str.~1, 85748 Garching, Germany}

\title{Electromagnetically Induced Transparency and Light Storage\\ in an Atomic Mott Insulator}

\begin{abstract}
We experimentally demonstrate electromagnetically
induced transparency and light storage with
ultracold $^{87}$Rb atoms in a Mott insulating
state in a three dimensional optical lattice. We
have observed light storage times of $\simeq
240$~ms, to our knowledge the longest ever
achieved in ultracold atomic samples. Using the
differential light shift caused by a spatially
inhomogeneous far detuned light field we imprint
a ``phase gradient" across the atomic sample,
resulting in controlled angular redirection of
the retrieved light pulse.
\end{abstract}
\date{March 1, 2009}




\maketitle Coherent interaction between light and
matter plays an important role in many quantum
information and quantum communication schemes
\cite{Duan01,Lukin03}. In particular, it is
desirable to transfer quantum states  from
photonic, ``flying" qubits to matter-based
systems for storage and processing
\cite{Masalas04}. In this context,
electromagnetically induced transparency (EIT)
has proven extremely useful, since it allows an
incoming light pulse to be converted into a
stationary superposition of internal states and
back into a light pulse
\cite{Fleischhauer03,Lukin00,Liu01,Philips01}.
This effect has successfully been used to map
quantum states of light onto cold atomic
ensembles \cite{Kimble08} or even to transmit
quantum information between two such remote
quantum memories \cite{Chaneliere05}. EIT and
light storage have been realized in crystals
\cite{Turukhin02}, atomic vapors
\cite{Kash99,Philips01} and in ultracold atomic
ensembles \cite{Hau99,Liu01,Ahufinger2002}. In
crystals, storage times of several seconds have
been achieved \cite{Longdell05}. In vapor cells,
inelastic collisions with other atoms or with the
walls usually limit the coherence times to a few
milliseconds \cite{Julsgaard04,Klein06}. In cold
atomic samples the light storage times are also
on a millisecond timescale \cite{Liu01}. Using
magnetically insensitive states, storage times of
up to 6~ms were recently observed even for single
quantum excitations in cold atomic gases, limited
by loss of atoms \cite{BZhao2009} or thermal
diffusion \cite{RZhao2009}.

Ultracold atoms in a Mott insulator (MI) state
with unity filling in a deep 3D optical lattice
are ideal for light storage, as they experience
no diffusion and no collisional interaction. In
the present work, we demonstrate EIT and long
light storage in such an environment. The minimal
dephasing observed allows for many possibilities
for processing stored information using advanced
manipulation techniques for atomic many-body
states in optical lattices (see
Ref.~\cite{BlochRMP} and references therein).
Light pulses can be stored in an atomic spin wave
in the MI, transformed, and then efficiently
mapped back into photonic modes. As an example of
such a spin-wave manipulation, we imprint a
``phase gradient" across the atomic sample using
a spatially varying differential light shift of
the two ground state levels. This spatial phase
gradient results in a controlled change of the
direction of the restored pulse. By controlling
non-classical atomic spin excitations, atoms in
optical lattices could even be turned into novel
non-classical light sources
\cite{Porras:2008,Pedersen:2008} or lead to
deterministic photonic phase gates at the single
photon level \cite{Masalas04}.

In our experiment we begin with ultracold
$^{87}$Rb atoms in the $\ket{F=1,
m_F=-1}\equiv\ket{1}$ state in an optical lattice
consisting of three mutually orthogonal
retroreflected laser beams each with
$1/e^2$-radius $\approx150~\mu$m. Two of the
lattice beams are red detuned
($\lambda_{y,z}=844$~nm), while the third is blue
detuned ($\lambda_x=765$~nm). In a sufficiently
deep lattice ($30~E_r$; $E_r=h^2\!/2m\lambda_z^2$
is the recoil energy), the many-body ground state
is a MI with a well-defined number of atoms on
each lattice site.

\begin{figure}[!t]
\begin{center}
  \includegraphics[width=7cm]{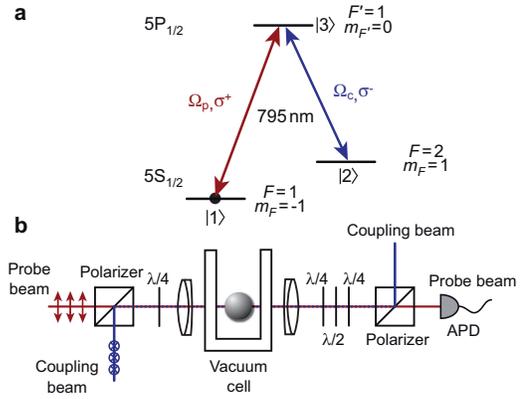}
 \caption{(a) EIT $\Lambda$-system in $^{87}$Rb (D$_1$-line).
 The transition between the
 two ground states $\ket{1}$ and $\ket{2}$ is
 insensitive to magnetic field fluctuations to
 first order at $B\simeq3.23$~G. (b) Experimental setup. Probe and
 coupling beam are used in a collinear
 configuration and are focussed onto the atoms in
the optical lattice.\label{fig:setup}}
\end{center}
\vspace{-0.5cm}
\end{figure}

For EIT, we use a $\Lambda$-system consisting of
the two Zeeman sublevels $\ket{1}$ and $\ket{F=2,
m_F=+1}\equiv\ket{2}$ of the 5S$_{1/2}$ ground
state, and the $\ket{F'=1, m_F=0}\equiv\ket{3}$
level of the 5P$_{1/2}$ excited state
[Fig.~\ref{fig:setup}(a)]. At a field of
$B\simeq3.23$~G, the states  $\ket{1}$ and
$\ket{2}$ have the same first-order Zeeman shift
\cite{Harber02}. The coupling laser light with
Rabi frequency $\Omega_c$ is $\sigma^-$-polarized
and is resonant with the
$\ket{2}\leftrightarrow\ket{3}$ transition. The
probe laser (Rabi frequency $\Omega_p$, frequency
$\omega_p$, $\sigma^+$-pol.) is phase locked to
the coupling laser with a difference frequency
corresponding to the ground state hyperfine
splitting. We use a collinear arrangement of
probe- and coupling beams
[Fig.~\ref{fig:setup}(b)] in order to avoid
momentum transfer to the atoms. The two beams are
overlapped on a Glan-Thompson polarizer before a
$\lambda/4$ waveplate converts the linear into
circular polarizations. A lens system focusses
the beams onto the atomic sample. The coupling
beam has a 1/e$^2$-radius of $\simeq150~\mu$m,
much larger than the diameter of the atomic
sample (typically $26~\mu$m), in order to
facilitate the alignment and to create a
spatially homogeneous coupling laser field. The
probe laser beam has a radius of $\simeq40~\mu$m.
The outgoing beams are separated using
polarization optics (suppression ratios of
$10^{3}-10^{4}$) and the probe beam is directed
onto an avalanche photodiode (APD, Analog Modules
712A-4).

\begin{figure}[t]
\begin{center}
  \includegraphics[width=7cm]{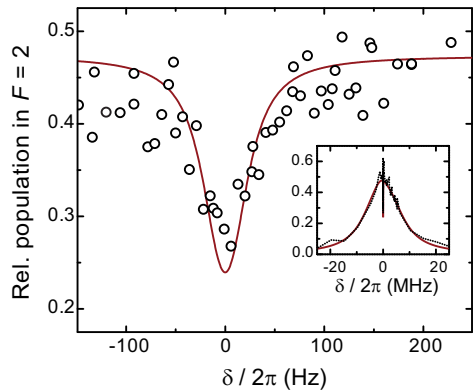}
\caption{Observation of EIT in a Mott insulator.
Shown is the fraction of atoms transferred from
$F=1$ to $F=2$ by a 200~ms probe laser pulse, as
a function of the two-photon detuning
$\delta=\omega_p-\omega_c-\omega_{21}$. The
observed EIT window has a width of $81(10)$~Hz.
The inset shows the total lineshape. The red line
is a prediction from a rate equation model (see
text).
}\label{fig:EIT}
\end{center}
\vspace{-0.5cm}
\end{figure}

We first observe EIT, in particular the existence
of a narrow transmission window, in an atomic
sample of $\simeq9\times10^4$ atoms, which in our
system corresponds to a MI with only singly
occupied sites. The atomic sample is an ellipsoid
with radii $r_x = 8.6~\mu$m and $r_{y,z} =
13.1~\mu$m. We shine in the coupling laser
($\Omega_c=2\pi\times26(5)$~kHz) and a weak probe
laser pulse ($\Omega_p=2\pi\times7(2)$~kHz) for
200~ms. Due to the small system size and low
powers necessary to achieve such a narrow EIT
window, a direct measurement of probe
transmission through the atom cloud is difficult
in our case. Instead, we measure the fraction of
atoms transferred by the probe laser to the $F=2$
manifold. We first detect the number of atoms
$N_{2}$ in $F=2$ by resonant absorption imaging.
A second image is taken $500~\mu$s later with a
repumper in order to also detect the atoms in the
$F=1$ manifold ($N=N_{2}+N_{1}$). The graphs in
Fig.~\ref{fig:EIT} show the relative population
transfer $N_{2}/N$ as a function of the
two-photon detuning. We observe an EIT
transmission window ($81(10)$~Hz FWHM) at the
center of the absorption line. We calculate the
fraction of atoms pumped from $\ket{1}$ into the
$F=2$ manifold by a rate equation model. It
includes the analytic expression for the linear
susceptibility given in
Ref.~\cite{Fleischhauer03} and also accounts for
the inhomogeneous optical depth (OD), which
arises from the ellipsoidal cloud shape. To
explain the observed population transfer to $F=2$
also at the center of the EIT window, we include
a decay rate of the $\ket{1}-\ket{2}$ coherence
$\gamma_{21}=2\pi\times10$~Hz and transfer to
$F=2$ by a fraction of $\pi$-polarized probe
laser light ($\Omega_p^\pi$) on the $\ket{1}
\rightarrow $\ket{F'=1, m_F=-1} transition. The
best agreement with the data is obtained with
$\Omega_c = 2\pi\times 27$~kHz,
$\Omega_p=2\pi\times3.9$~kHz and
$\Omega_p^\pi=0.2\,\Omega_p$ (red line in
Fig.~\ref{fig:EIT}), which are close to the
measured values.

\begin{figure}[b]
\begin{center}
  \includegraphics[width=7cm]{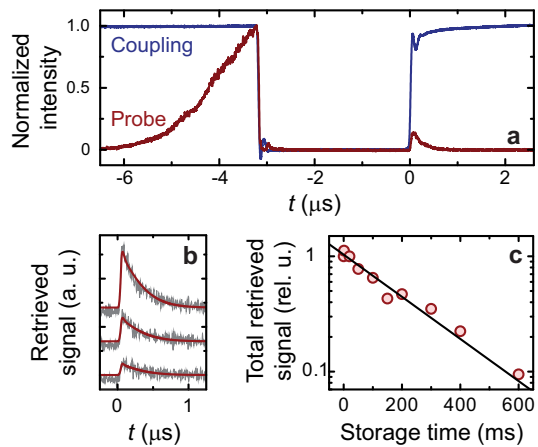}
\end{center}
\vspace{-0.5cm} \caption{Light storage. (a)
Intensity of coupling (probe) beams, recorded on
a photodiode before (after) the atomic sample
during a light storage experiment with 3~$\mu$s
storage time in a thermal cloud ($\Omega_c =
2\pi\times 4.9$~MHz, $\Omega_{p}=2\pi\times
920$~kHz, $\simeq10^6$ atoms) (b) Retrieved
pulses for storage times of $t_S=1$~ms, 200~ms,
400~ms (from top to bottom) in a Mott insulator.
Traces are offset for better visibility
($\Omega_c= 2\pi\times 4.5$~MHz,
$\Omega_{p}=2\pi\times 1.5$~MHz,
$\simeq9\times10^4$ atoms). (c) Retrieved signal
(relative to the signal at $t_S=3~\mu$s) as a
function of $t_S$. The line is an exponential fit
with decay time $\tau=238(20)$~ms.}
\label{fig:lightStorage}
\end{figure}
As a second experiment, we demonstrate the
storage of light pulses (Fig.~3). After turning
on the coupling beam, we apply a Gaussian-shaped
probe pulse with 2.8~$\mu$s FWHM. At the peak of
this pulse, we shut off the probe and coupling
beams simultaneously, within less than 50~ns.
After waiting for a variable storage time, we
turn on the coupling beam again and monitor the
restored probe pulse on an APD. The second,
retrieved pulse is much smaller than the first,
incident pulse. From the ratio of their areas, we
estimate the storage efficiency to be 3\% for a
large thermal cloud [Fig.~3(a)] and 0.3\% for the
MI. The small efficiency is partly caused by the
mismatch of the size of the probe beam and the
atomic sample (18\% geometrical overlap for the
MI), leakage of the probe beam due to the finite
OD of the sample (peak OD $\alpha=6.3$, see
definition in Ref.~\cite{Lukin03}), and due to
spontaneous emission during writing and retrieval
phases. From numerical simulations based on the
equations in \cite{Phillips08}, we estimate the
efficiencies due to leakage and spontaneous
emission as 11\% for short storage times
($3~\mu$s). The same simulations were used to
reproduce the retrieved pulse shapes shown in
Fig.~3(b) with no free parameters other than the
amplitude. Not included in the simulations are
effects due to imperfect polarizations of probe and
coupling beams.

We use the energy (integrated intensity) of the
restored pulse as a measure of the stored light
signal. As shown in
Fig.~\ref{fig:lightStorage}(c), fitting an
exponential decay to the retrieved pulse power as
a function of storage time yields a decay time
constant of $\tau=238(20)$~ms. To independently
measure the coherence time of the
$\ket{1}+\ket{2}$ superposition, we performed a
Ramsey experiment on the same states using an
rf+microwave two-photon transition
\cite{Harber02}. The visibility of the Ramsey
fringes decays with a time constant of $2
\tau=436(22)$ ms, indicating that the decay of
the stored light pulse is not caused by residual
coupling light present during the storage time.
The factor of two arises since the Ramsey fringe
contrast measures the decay of the quantum
amplitude coherence \cite{Widera2002}, whereas in
the EIT signal we measure an intensity. A
$\pi$-echo pulse does not restore the Ramsey
signal contrast, so the decay time has to be
attributed to an irreversible dephasing
mechanism. We have ruled out magnetic field noise
by measuring coherence times away from the
``magic" field at 3.23~G. The coherence times are
nearly unchanged at 6~G and 2~G, where the
differential shift of the
$\ket{1}\leftrightarrow\ket{2}$ transition is at
least an order of magnitude more sensitive to
magnetic field fluctuations. We measured the
coherence time vs lattice depth and found a
maximum at $30-40~E_r$ for our experimental
parameters. This indicates that the source of the
coherence decay is due to heating in the optical
lattice and to finite tunneling. The latter leads
to an increased probability of having more than
one atom per lattice site. In this case the
interaction energy in the doubly occupied sites
leads to an onsite dephasing with respect to the
singly occupied sites. Increasing the lattice
depth improves the coherence times due to the
suppression of tunneling, but in turn the heating
due to spontaneous light scattering and technical
noise increases. Our analysis suggests that by
simply using farther detuned lattices, even
longer light storage times can be achieved.

In a non-collinear geometry, the difference in
the wavevectors of coupling and probe beams,
${\bf k}_{c}-{\bf  k}_{p}$, is stored as a
spatial gradient in the phase of the atomic
superposition state \cite{Liu01}:
\begin{equation}
 \ket{D}
 =
 \frac{\Omega_{c}\ket{1} - \Omega_{p}e^{i({\bf
 k}_{c}-{\bf  k}_{p}){\bf r}} \ket{2}}
 {\sqrt{\Omega_{c}^2+\Omega_{p}^2} }.
 \label{eq:D}
\end{equation}

Here we reverse the logic leading to
Eq.~(\ref{eq:D}), and show that imprinting a
phase gradient on a stored-light state can change
the direction of the restored pulse. This is
similar to the work demonstrating deflection of
light in a vapor cell by a magnetic gradient
field \cite{Weitz06} or by an inhomogeneous laser
beam \cite{Sautenkov07}. In our experiment, we
first store a pulse in a MI with
$\simeq2.5\times10^5$ atoms, and a lattice depth
of $30~E_{r}$  using the sequence described above
($\Omega_c=2\pi\times4.3$~MHz,
$\Omega_p=2\pi\times3.8$~MHz).
\begin{figure}[!t]
\begin{center}
  \includegraphics[width=7cm]{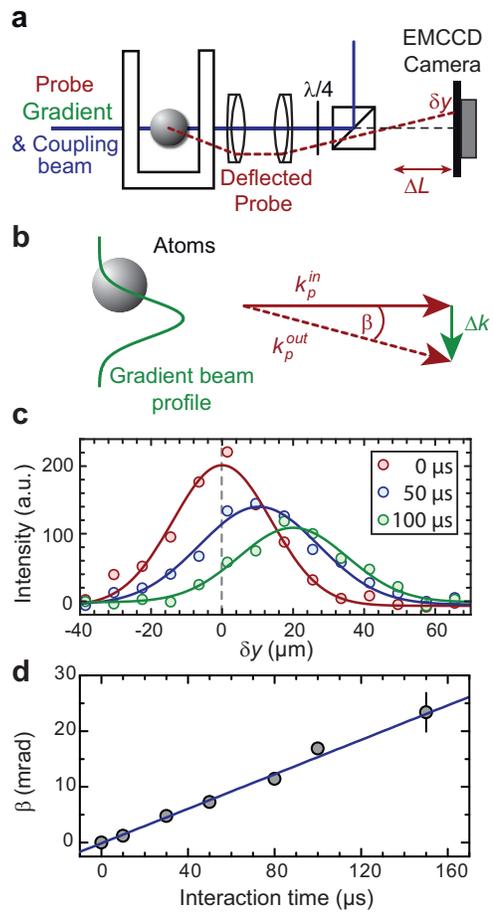}
\end{center}
\caption{Angular deflection of a stored light
pulse. (a) The deflected light pulse is detected
with an EMCCD camera. (b) A detuned laser beam
with a spatially varying intensity profile across
the atoms creates a spatial phase gradient via
the differential light shift. (c) Row sums of the
CCD images from the deflected light pulse for
different interaction times. Each curve is
averaged over 5 runs and the background due to
the coupling beam is subtracted. (d) Deflection
angle $\beta$ as a function of the interaction
time $t_{int}$ with the gradient beam. The blue
line is a linear fit. } \label{fig:Deflection}
\end{figure}
Before retrieving the pulse after 10~ms storage
time, we shine in an additional $\sigma^+$
polarized laser ($w_0=42(8)~\mu$m), aligned
$20(5)~\mu$m away from the center of the atomic
cloud [Fig.~\ref{fig:Deflection}(b)]. The laser
is red detuned from the
$\ket{2}$$\leftrightarrow$$\ket{3}$ transition by
$-20$~GHz, which causes spatially inhomogeneous
light shifts $\Delta_{1,2}(\vec r)$ of the two
ground state levels due to the Gaussian intensity
profile. Shining in this laser for an interaction
time $t_{int}$ induces a local dephasing between
$\ket{1}$ and $\ket{2}$ of $ \phi(\vec r)=
[\Delta_1(\vec r)-\Delta_2(\vec
r)]t_{int}/\hbar$. In our experiment, the
maxi\-mum laser intensity is $2.3(2)$~W/cm$^2$,
which produces a differential light shift of
$\Delta_1-\Delta_2 = 2\pi\times7.7$~kHz at the
center of the atomic cloud. The interaction time
$t_{int}$ is varied from $0$ to $150~\mu$s. The
deflection angle is
\begin{equation}
 \beta \simeq \frac{\Delta k}{k_{p}}
 \quad \mbox{with}\quad
 \Delta k =
 \frac{d(\Delta_1-\Delta_2)}{dy}
  \frac{t_{int}}{\hbar},
\end{equation}
where $k_{p}=2\pi/\lambda$ is the wavevector of
the probe laser beam.

The deflected pulse is detected using an electron
multiplying CCD (EMCCD, ANDOR iXon DV885), see
Fig.~\ref{fig:Deflection}(a). In order to reveal
the deflection, the camera is placed out of the
focal plane by translating the last lens before
the camera. The detected signal on the EMCCD
camera for $t_{int}=0~\mu$s contains about
$1.1\times10^5$ counts (corresponding to
$3.4\times10^3$ photons). This signal was then
summed along the $z$-direction and averaged over
$5-12$ runs for better visibility. To each of
these integrated pulses we fit a 1D Gaussian and
determine the position shift $\delta y$ of the
deflected beam. From $\delta y$ and the camera
position with respect to the focal plane, we
determine the deflection angle $\beta$. The
result is summarized in
Fig.~\ref{fig:Deflection}(d) together with a
linear fit. The fitted slope $d\beta/d
t_{int}=155(5)~\mu$rad/$\mu$s is close to the
value of $232(46)~\mu$rad/$\mu$s calculated from
our experimental parameters. The error takes into
account the uncertainties of the gradient beam
power, waist and the alignment.

In summary we have demonstrated EIT, light
storage and retrieval from an atomic Mott
insulator. We have observed very long storage
times of about 240~ms, where the storage time is
limited by heating from the lattice and by
tunneling. We also demonstrated that a stored
pulse can be controlled and redirected by
imprinting a spatial phase gradient with a laser
beam.

In the future, it would be interesting to extend
this technique to more complex light fields in
order to process and manipulate information
stored in spin structures, which can then be
analyzed by measuring the direction and shape of
the retrieved pulse. In contrast two the usual
manipulation of the spins by microwave radiation,
EIT also allows the imprinting of elaborate phase
structures generated by holograms such as images
or vortices \cite{Davidson07}. This could
facilitate the study of far-from equilibrium
spinor gases, or allow the storage of a doubly
charged $m=2$ vortex in the MI phase, where it is
expected to be stable in contrast to a BEC
\cite{Shin2004}. Another interesting prospect is
to use the MI as a genuine quantum memory to
store and to retrieve single photons
\cite{BZhao2009,RZhao2009}. By using an optimized
geometry with a higher OD, storage of an entire
pulse or pulse sequence can be achieved. As an
alternative to storing light pulses, one can also
directly create an atomic superposition. Turning
on the coupling field then leads to the creation
of a probe field. We are currently exploring the
use of such a created light pulse as a novel
probe for classical or entangled atomic spin
states in an optical lattice. Ultimately, the
generation of such non-classical spin states and
the direct mapping onto photonic states could
lead to a new generation of non-classical light
sources \cite{Porras:2008,Pedersen:2008}.

We acknowledge support by DIP, DFG, EU (IP
SCALA), AFOSR, and DARPA (OLE) and the Fulbright
Association (J.D.T.). We thank A. V. Gorshkov for
helpful discussions.

\end{document}